\documentstyle{europhys}

\def\And{{\rm and\ }}

\newif\ifboo \boofalse

\input epsf

\begin{document}

\euro{x}{x}{x}{x}
\Date{May 14 1998}
\shorttitle{MAGNETIC FARADAY--INSTABILITY}

\title{Magnetic Faraday--Instability}

\author
{
        T. Mahr\inst{1}\footnote{email:
        thomas.mahr@physik.uni-magdeburg.de} \And
        I. Rehberg\inst{1}\footnote{email:
        ingo.rehberg@physik.uni-magdeburg.de}
}

\institute
{
     \inst{1}   Institut f\"ur Experimentelle Physik,
                Otto-von-Guericke-Universit\"at,
                Postfach 4120,
                D-39016 Magdeburg,
                Germany
}

\rec{}{}

\pacs
{
        \Pacs{47}{20.-k}{Hydrodynamic stability}
        \Pacs{47}{35.+i}{Hydrodynamic waves}
        \Pacs{75}{50.Mm}{Magnetic liquids}
}

\maketitle

\begin{abstract}

In a magnetic fluid parametrically driven surface waves can be excited by
an external oscillating magnetic field. A static magnetic field changes
the restoring forces and damping coefficients of the various surface
waves. This property enables the excitation of both subharmonic and
harmonic responses of the standing waves. 

\end{abstract}

\section{Introduction}

``Most fluids, if not all, may be used to produce these crispations, but
some with particular advantages; alcohol, oil of turpentine, white of egg,
ink and milk produce them''. What Michael Faraday describes here is the
parametric excitation of surface waves, nowadays known as the
Faraday-Instability \cite{Faraday}. Today it is a popular experimental
system for the study of parametrically excited instabilities, pattern
formation and spatio-temporal chaos \cite{CrossHohenberg}. Following
Faraday's suggestion we use a fluid which was not available to him: a
superparamagnetic fluid, more popularly known as ferrofluid or magnetic
fluid. Its particular advantage is that the properties of the fluid can be
tuned in a wide range by an external magnetic field, and that Faraday's
parametric excitation can be achieved by temporal modulation of that
magnetic field, i.e. without a mechanical driving. The mechanism of the
parametric driving is however very similar to the mechanic case: the
spontaneously created surface deformation of a standing wave is amplified
by the concentration of the magnetic field in the neighborhood of a wave
crest (demagnetization). This amplification takes place twice during the
oscillation cycle of the standing wave: peaks which are half a wavelength
apart are amplified at time differences of half a driving period $T_D$.

The idea to excite surface waves magnetically in a ferrofluid is fairly
similar to the electric excitation of dielectric fluids \cite{Briskman68}
and thus not new \cite{Okubo,Bashtovoi}. In those experiments a magnetic
field perpendicular to the surface of the fluid was used. Related to this
idea is the excitation of surface waves with a horizontal magnetic field,
which however leads to oscillations not described by the Mathieu-equation
\cite{BacriCebers94}. In the experiment presented in this paper, we use a
steady magnetic field to tune the fluid parameters, and an oscillating
field perpendicular to the surface to produce a parametric excitation of
standing surface waves. With this magnetic excitation we are able to
obtain the classical subharmonic response, and to measure its threshold
curve. More importantly, we succeed in measuring for the first time
threshold values for higher resonance tongues, which are not easily
accessible for ordinary liquids \cite{HWMueller}. This is possible because
magnetic fluids have a particular advantage: They allow for a tuning of
the restoring force of the surface waves by means of a steady external
magnetic field. In particular, surface waves with the critical wave number
of the Rosensweig--Instability \cite{RosenBuch} have zero frequency and
are thus undamped at the critcal field $H_c$ for the onset of the
instability.

\section{Experimental Setup}

\begin{figure}[h]
\epsfxsize=15cm
\epsfbox{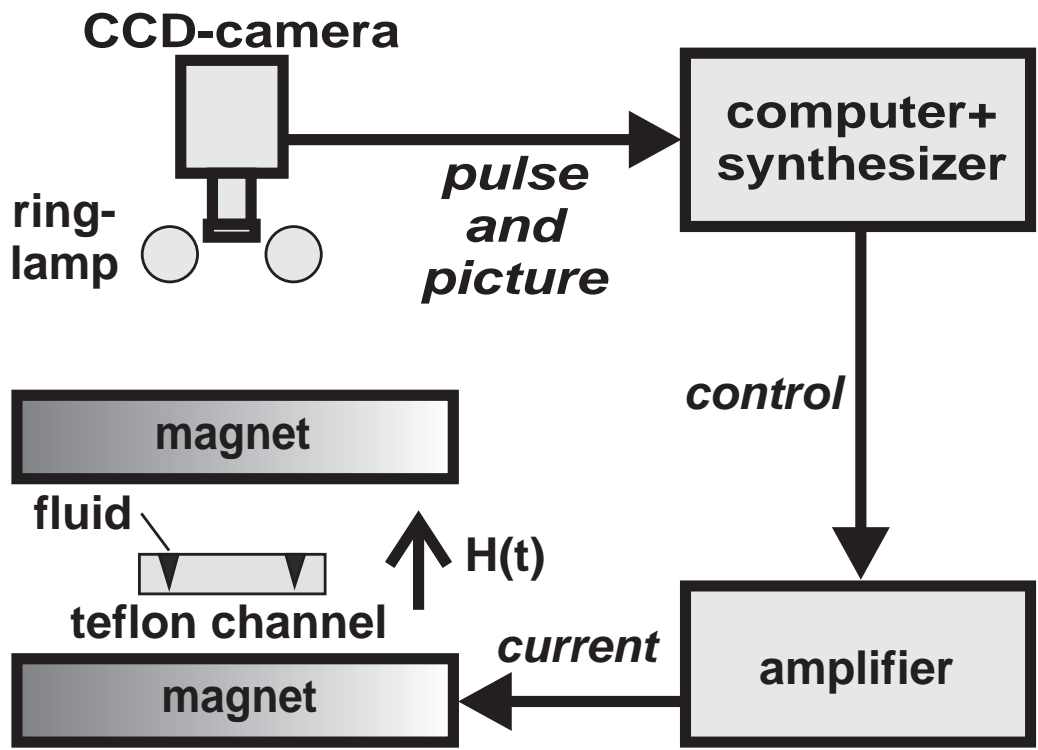}
\caption {\label{setup} Experimental setup. }
\end{figure}

The experimental setup is shown in Fig.~\ref{setup}. The magnetic fluids
are poured into a V-shaped circular Teflon channel of 60 mm diameter
\cite{MaRe95}. The cylindrical geometry prevents disturbances by meniscus
induced surface waves, and allows an evaluation of the wavennumber by
means of a Fast Fourier Transformation, FFT \cite{NumRec}. The channel has
a depth of 5 mm and upper width of 4 mm which is smaller than the typical
wavelengths in the experiment. The upper part of the channel has a slope
of $15^{\circ}$ in order to damp out surface waves in radial direction. 

We use the commercially available magnetic fluids EMG 705 and EMG 909
(Ferrofluidics). The properties of the water-based fluid EMG 705 are:
density $\rho = 1190$ kgm$^{-3}$, surface tension $\sigma = 4.75 \cdot
10^{-2}$ kgs$^{-2}$, initial magnetic permeability $\mu = 1.56$, magnetic
saturation $ M_S = 1.6 \cdot 10^4$ Am$^{-1}$, dynamic viscosity $ \eta= 5
\cdot 10^{-3}$ Nsm$^{-2}$, yielding a critical field for the onset of the
Rosensweig-Instability $H_C = 9.9 \cdot 10^3$ Am$^{-1}$. The properties of
the oil-based fluid EMG 909 are: density $\rho = 1020$ kgm$^{-3}$, surface
tension $\sigma = 2.65 \cdot 10^{-2}$ kgs$^{-2}$, initial magnetic
permeability $\mu = 1.8$, magnetic saturation $ M_S = 1.6 \cdot 10^4$
Am$^{-1}$, dynamic viscosity $ \eta= 6 \cdot 10^{-3}$ Nsm$^{-2}$, and the
critical field for the onset of the Rosensweig-Instability is $H_C = 7.9
\cdot 10^3$ Am$^{-1}$.

The channel is placed in the center of a pair of Helmholtz-coils
(Oswald-Magnetfeldtechnik), with an inner diameter of 40 cm. One coil
consist of 474 windings of flat copper wire with a width of 4.5 mm and
thickness 2.5 mm. A current of about 5 A is then sufficient to produce the
magnetic field of about $8 \cdot 10^3$ Am$^{-1}$ used in this experiment.
The static field is monitored by means of a hall probe (Group 3 DTM-141
Digital Teslameter) located near the surface of the channel.

The light from a tungsten bulb, placed in the center of the top of the
channel, is reflected at the surface of the fluid and directed towards the
CCD--camera placed 80 cm above the center of the channel. The camera
(Philips LDH 0600/00) works in the interlaced mode at 50 Hz using an
exposure time of 40 ms.

The analysis of the images and the control of the experiment is done with
a 90 MHz Pentium--PC, equipped with a $512 \times 512$ 8--bit frame
grabber (Data Translation DT2853), a programmable counter (8253) located
on a multifunction I/O--board (Meilhaus ME--30), and a synthesizer--board
(WSB--10). The counters are used to keep track of the pacing--frequency of
the synthesizer--board. Their output is used to trigger the camera in any
desired phase of the driving oscillation of the magnetic field. This
phase--locked technique between the driving and the sampling ensures a
jitter--free measurement of the amplitude. By keeping track of the
synthesizer pace the computer moreover manages the writing of the data
into the synthesizer memory at times when no conflict with the
DA--converter arises. This allows for smooth switching of the amplitude of
the AC--component of the magnetic field \cite{MaRe96}. 

The wave--signal is amplified by a linear amplifier (fug NLN 5200 M--260).
The resulting driving magnetic field is $H(t) = H_0 + \Delta H \sin{2\pi t
f_D}$, with $H_0$ as the static and $\Delta H$ as the oscillating part of
the magnetic field; $f_D$ is the driving frequency and $H_0>\Delta H$.

\section{Experimental Results}

\begin{figure}[h]
\hspace{1.5cm}
\epsfxsize=12cm
\epsfbox{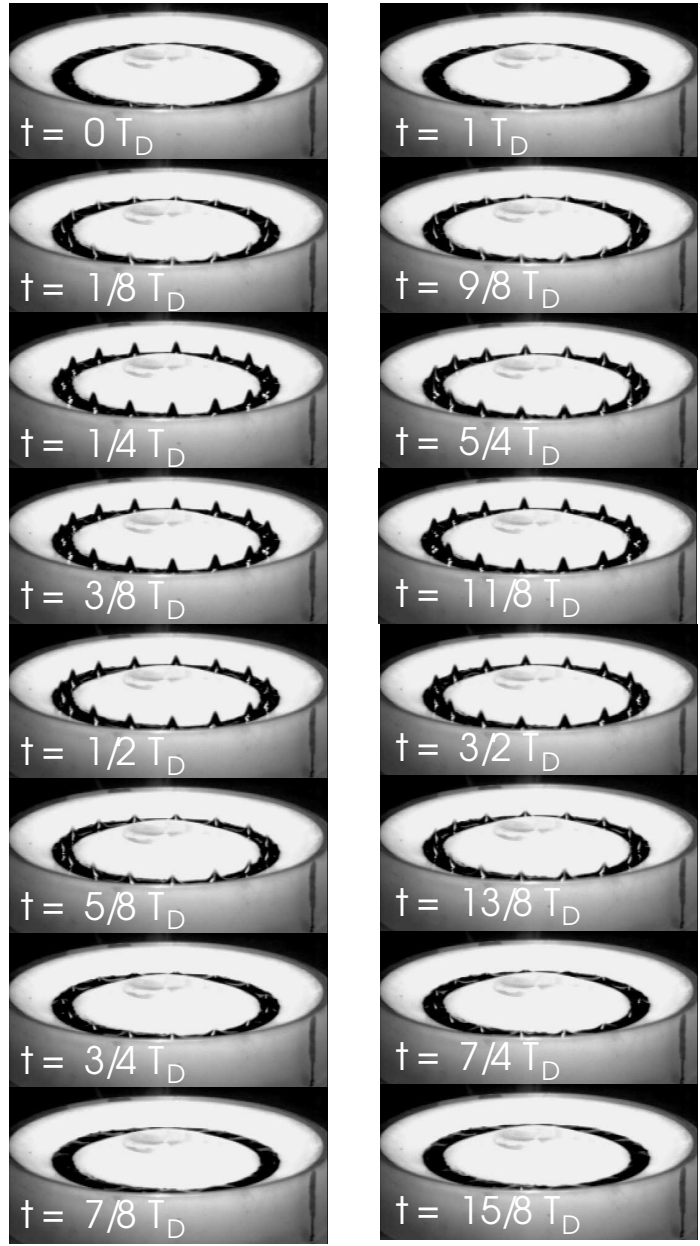}
\caption {\label{movie} 16 snapshots of a parametrically excited
standing wave at $T_D$ = 0.1 $s$, $H_0$ = 0.95 $H_C$, $\Delta H$ =
0.21 $H_C$, $H_C = 6.8 \cdot 10^3 Am^{-1}$ during a time of 2 $T_D$.
The phases are indicated in the left--bottom corner of each picture.
The fluid is EMG 909. }

\end{figure}

Using only a static magnetic field, $H_0$, leads to the
Rosensweig--Instability at the threshold $H_C$. In Fig.~\ref{movie} we
demonstrate the existence of parametrically excited surface waves for
$H_0<H_C$: when increasing the oscillating part, $\Delta H$, of the
driving above a certain threshold the flat surface of the magnetic fluid
becomes unstable and a standing wave ensues, as illustrated by a series of
snapshots of the behavior of the surface of the fluid during the time
interval $t=2T_D$. Here the oil--based fluid EMG 909 is excited at
$f_D=10$ Hz, $H_0=0.95$ $H_C$ and $\Delta H = 0.21$ $H_C$. Each of the 16
pictures in Fig.~\ref{movie} corresponds to a certain phase which is
labeled in the left--bottom corner of each picture. The first driving
period, $T_D$, is seen in the left column, the following second period in
the right. Starting at a nearly flat surface at $t=0$ wave crests with the
wavelength $\lambda$ grow to maximum amplitude at $t=3/8T_D$ and decrease
to a flat surface at $t=7/8 T_D$. At the next driving period the wave
crests grow to a maximum amplitude at $t=11/8 T_D$ with a $\lambda/2$
shift of the position of the peaks with respect to the preceding period.
This clearly indicates that we observe a standing wave with the response
period $T_{resp} = 2$ $T_D$. 

\begin{figure}[h]
\epsfxsize=15cm
\epsfbox{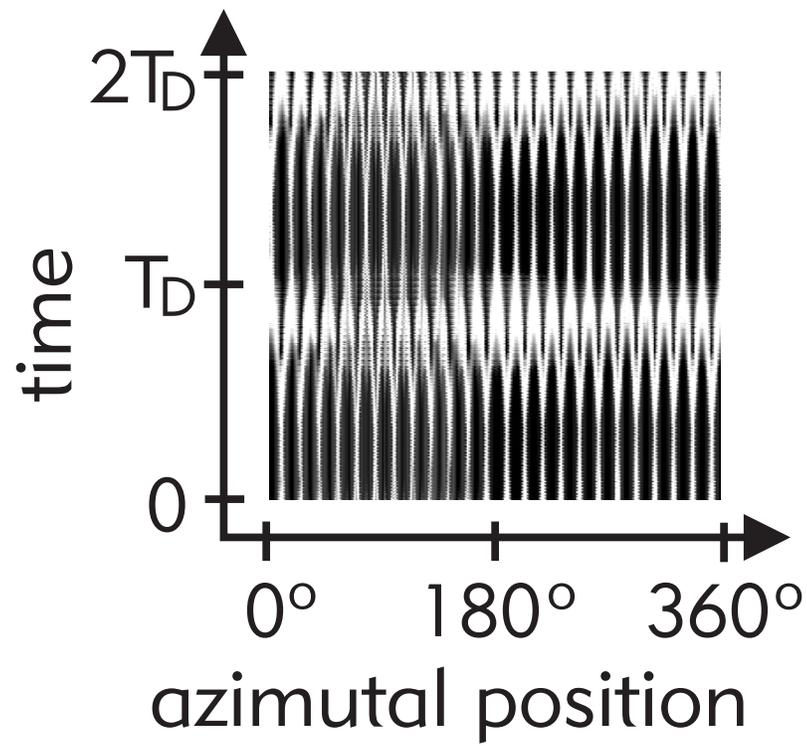}
\caption {\label{strob} Space--time plot for $f_D$ = 15 Hz, $H_0$ =
0.98 $H_C$, $\Delta H$ = 0.24 $H_C$, $H_C = 8.6 \cdot 10^3 Am^{-1}$.
The fluid is EMG 909.}
\end{figure}

For a more detailed characterization of the spatio--temporal behavior we
measure the light intensity reflected from the surface along the annulus
as described in detail in \cite{MaRe95}. The resulting spatio--temporal
representation is shown in Fig.~\ref{strob}. The dark zones correspond to
wave troughs and a high light intensity represents crests of the standing
waves. A time interval of two driving periods is shown and the standing
wave character with a period of $2 T_D$ is clearly visible.

\begin{figure}[h]
\epsfxsize=15cm
\epsfbox{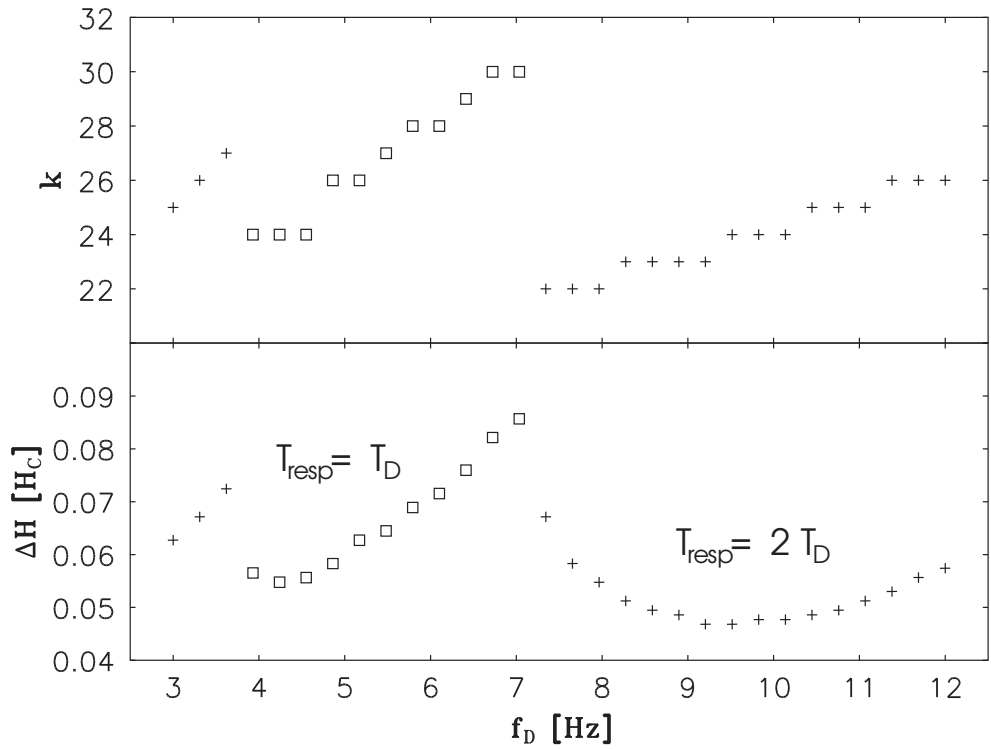}
\caption {\label{oil} At a fixed value of $H_0$ = 0.97 $H_C$ for
each value of the driving frequency $f_D$ the oscillating part is
increased until the flat surface becomes unstable. While in the lower
diagram these thresholds are indicated, in the upper diagram the
wave numbers of the resulting standing wave are shown. The step
$\Delta k = \mp 1$ corresponds to the destruction or creation of one
wavelength. Squares mark the $1T_D$--states and crosses the
$2T_D$--states. The fluid is EMG 909. $H_C = 7.9 \cdot 10^3 Am^{-1}$,
$k_C$ = 25.}
\end{figure}

In the following, we examine the stability of the flat surface in
dependence of the driving frequencies, $f_D$, and we present measurements
of the threshold for the onset of the Faraday--Instability. To do so, we
fix the static part of the magnetic field and the driving frequency. The
oscillation amplitude, $\Delta H$, is increased until the flat surface
becomes unstable. The criterion for instability is obtained from the
FFT--analysis of the light intensity measured along the channel: after
switching to a new amplitude we wait $15 s$, if then the intensity of the
peak in the power spectrum is larger than its mean value by more than a
factor of five we conclude that the instability threshold has been
crossed. We pinpoint the threshold by means of binary interval search
algorithm with an accuracy corresponding to one bit of the 12--bit
synthesizer card, which corresponds to an accuracy better than one
percent. The advantage of this procedure is that it works both for sub--
and supercritical bifurcations of the surface instability. These threshold
values are plotted for the oil based fluid EMG 909 at $H_0=0.97$ $H_C$ as
a function of the driving frequency in the lower part of Fig.~\ref{oil}. 

This clearly shows that the response with 2 $T_D$ is dominant, a fact
which can be explained on the basis of the theory presented in Fig.~4 of
Ref.~\cite{Bashtovoi}. According to that theory, the half frequency regime
is indeed the dominant one, a behavior which is typical for the parametric
driving of surface waves.

In addition the FFT yields the wave number $k$ of the surface wave which
is presented in the upper part of Fig.~\ref{oil}. This plot clearly
indicates wave number jumps along the subharmonic and harmonic branches.
The step $\Delta k = \mp 1$ corresponds to the destruction or creation of
one wavelength. In principal one would expect to observe minima of the
critical $\Delta H$ for a fixed value at $k$, as shown in Fig.~20 of
Ref.~\cite{Douady90}. The resolution in $\Delta H$ of our apparatus is not
sufficiently fine for this purpose. Moreover a spatio--temporal spectral
analysis allows for discrimination between harmonic and subharmonic
responses: for frequencies $f_D>7$ Hz the surface responds with $T_{resp}
= 2 T_D$. 

\begin{figure}[h]
\epsfxsize=15cm
\epsfbox{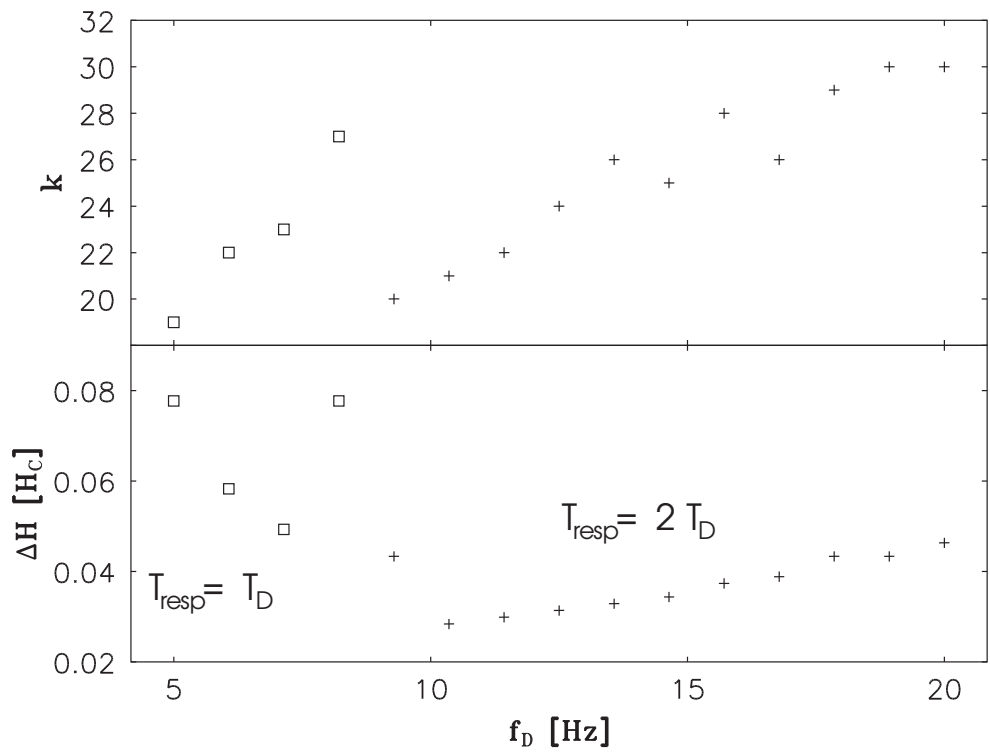}
\caption {\label{water} At a fixed value of $H_0$ = 0.93 $H_C$ for
each value of the driving frequency $f_D$ the oscillating part is
increased until the flat surface becomes unstable. While in the lower
diagram these thresholds are indicated, in the upper diagram the
wave numbers of the resulting standing wave are shown. Squares mark
the $1T_D$--states and crosses the $2T_D$--states. The fluid is EMG
705. $H_C = 9.9 \cdot 10^3 Am^{-1}$, $k_C$ = 20.}
\end{figure}

We have performed similar measurements with a less viscous fluid, namely
the water--based EMG 705. The results are presented in Fig.~\ref{water}.
We find the two modes $T_{resp} = T_D$ and $T_{resp} = 2 T_D$. Note that
around the minimum at $f_D = 10.4$ Hz of the subharmonic tongue, i.~e.~the
line in the $f_D$--$\Delta
$H plane where the instability sets in with a frequency of $2 T_D$,
the total magnetic field $|H_0|+|\Delta H|$ is well below the critical
value $H_C$ of the Rosensweig--Instability. This means that the parametric
driving is more efficient in this fluid, which is easily explained by its
smaller viscosity.

The threshold of the amplitude $\Delta H$ in Fig.~\ref{oil} shows a
minimum at about $f_D=9.3$ Hz. For even lower frequencies the harmonic
response has a lower stability threshold than the subharmonic response. To
some extend, this might be due to the fact that for increasing wavelengths
the dissipation in the bottom layer is enhanced \cite{HWMueller}. More
importantly, this might be a manifestation of the advent of the
Rosensweig--Instability which occurs at about $k_c$=25 This damping favors
the higher resonance tongue with $T_{resp}=T_D$ which has a larger wave
number $k$. The harmonic resonance tongue is observed in the range between
3.6 Hz $< f_D <$ 7.0 Hz. For driving frequencies lower than 3.6 Hz the
first instability is again the subharmonic response with $T_{resp} = 2
T_D$. The qualitative behavior of the threshold graphs $\Delta H(f_D)$ and
$k(f_D)$ are similar to the numerically obtained linear threshold
amplitude and critical wave number shown in \cite{CerdaTirapegui}.
Nevertheless, there is a striking difference when comparing our data in
Fig.~\ref{oil} with Fig.~2 of Ref.~\cite{CerdaTirapegui}: the linear
theory for mechanically excited parametric resonance predicts $T_{resp} =
2/3 T_D$, while we measure a response period $T_{resp} = 2 T_D$. We
believe that this is due to a nonlinear effect, the $2/3 T_D$--mode is
presumably not stable. Our measurement procedure only allows for a
determination of the linear threshold, while the nonlinear aspects of
magnetically driven oscillations are many and various and are 
presented elsewhere \cite{MaRe96}.

\section{Summary and Conclusion}

In summary, we have obtained parametrically driven higher resonance
tongues by means of tuning the viscous damping of a magnetic fluid with an
external magnetic field. This is made possible by a specific peculiarity
of magnetic fluids, namely a very weak damping of surface waves when the
wave number is close to the critical wave number of the
Rosensweig--Instability. One might expect that the interaction between the
$2T$ with the $1T$ resonance leads to complicated spatial patterns, a
supposition which is currently under investigation.  

\section{Acknowledgment}

The experiments are supported by the 'Deutsche Forschungsgemeinschaft'
through Re588/10. It is a pleasure to thank H.~W.~M\"uller and R.~Richter
for helpful discussions.


\end{document}